\documentclass[pdflatex,sn-mathphys-num]{sn-jnl}

\usepackage{graphicx}%
\usepackage{multirow}%
\usepackage{amsmath,amssymb,amsfonts}%
\usepackage{amsthm}%
\usepackage{mathrsfs}%
\usepackage[title]{appendix}%
\usepackage{xcolor}%
\usepackage{textcomp}%
\usepackage{manyfoot}%
\usepackage{booktabs}%
\usepackage{algorithm}%
\usepackage{algorithmicx}%
\usepackage{algpseudocode}%
\usepackage{listings}%

\usepackage{orcidlink}
\usepackage{caption}
\usepackage{subcaption}
\usepackage{url}
\usepackage{hyperref}

\graphicspath{ {./imgs/} }

\theoremstyle{thmstyleone}%
\theoremstyle{thmstyletwo}%

\theoremstyle{thmstylethree}%

\begin{document}

\title[Article Title]{PD-L1 Classification of Weakly-Labeled Whole Slide Images of Breast Cancer}


\author*[1]{\fnm{Giacomo} \sur{Cignoni}}\email{g.cignoni3@studenti.unipi.it \orcidlink{0000-0002-9138-0891}}

\author[2]{\fnm{Cristian} \sur{Scatena}}\email{cristian.scatena@unipi.it \orcidlink{0000-0002-4862-0845}}

\author[3]{\fnm{Chiara} \sur{Frascarelli}}\email{chiara.frascarelli@ieo.it}

\author[3]{\fnm{Nicola} \sur{Fusco}}\email{nicola.fusco@ieo.it \orcidlink{0000-0002-9101-9131}}

\author[2]{\fnm{Antonio Giuseppe} \sur{Naccarato}}\email{giuseppe.naccarato@unipi.it \orcidlink{0000-0002-8195-9445}}

\author[2]{\fnm{Giuseppe Nicoló} \sur{Fanelli}}\email{nicolo.fanelli@unipi.it \orcidlink{0000-0001-7069-7980}}
\equalcont{Co-Last authors.}

\author[1]{\fnm{Alina} \sur{S\^irbu}}\email{alina.sirbu@unipi.it \orcidlink{0000-0002-3947-7143}}
\equalcont{Co-Last authors.}

\affil*[1]{\orgdiv{Department of Computer Science}, \orgname{University of Pisa}
}

\affil[2]{\orgdiv{Department of Translational Research and New Technologies in Medicine and Surgery}, \orgname{University of Pisa}
}

\affil[3]{\orgdiv{Division of Pathology}, \orgname{IEO (European Institute of Oncology), IRCCS}
}

\abstract{Specific and effective breast cancer therapy relies on the accurate quantification of PD-L1 positivity in tumors, which appears in the form of brown stainings in high resolution whole slide images (WSIs). However, the retrieval and extensive labeling of PD-L1 stained WSIs is a time-consuming and challenging task for pathologists, resulting in low reproducibility, especially for borderline images. This study aims to develop and compare models able to classify PD-L1 positivity of breast cancer samples based on WSI analysis, relying only on WSI-level labels. The task consists of two phases: identifying regions of interest (ROI) and classifying tumors as PD-L1 positive or negative. For the latter, two model categories were developed, with different feature extraction methodologies. The first encodes images based on the colour distance from a base color. The second uses a convolutional autoencoder to obtain embeddings of WSI tiles, and aggregates them into a WSI-level embedding. For both model types, features are fed into downstream ML classifiers. Two datasets from different clinical centers were used in two different training configurations: (1) training on one dataset and testing on the other; (2) combining the datasets.
We also tested the performance with or without human preprocessing to remove brown artefacts
Colour distance based models achieve the best performances on testing configuration (1) with artefact removal, while autoencoder-based models are superior in the remaining cases, which are prone to greater data variability.}



\maketitle

\section{Introduction}\label{sec1}

Breast cancer (BC), the most frequent malignancy in women, is a heterogeneous disease from the clinical, histological, and molecular viewpoint. 
Triple negative breast cancer (TNBC) has a more aggressive clinical course (about 15–20\% of BCs). 
PD-L1 expression is used to evaluate metastatic TNBC, where the administration of the drug atezolizumab has been recently approved \cite{pdl1_pathway}. PD-L1 is a transmembrane protein that downregulates antitumor immune responses. 
The evaluation consists in immunohistochemical (IHC) staining with Roche VENTANA PD-L1 (SP142) Assay~\cite{ventana}. SP142 stains tumor-infiltrating immune cells (IC) present in the intratumoral and contiguous peritumoral stroma; stained cells are recognizable by the brown tinted staining. If the specimen contains PD-L1 staining of any intensity in IC occupying $\geq$ 1\% of tumor area, the case will be assigned as \emph{PD-L1 positive}. 

Currently, the evaluation of the amount of staining is performed visually by trained histo-pathologists. The low threshold (1\%) and the presence of artifacts (errors in the staining process, such as blank spots, or DAB brown spots) makes evaluation challenging and reduces the assay's standardization. Therefore, digital evaluation on whole slide images (WSIs) is potentially an innovative approach able to reduce the inter-observer variability \cite{wsi_imaging,intro_digital_img_analysis}.

Digital image analysis is widely used in medicine for diagnosis and prognosis, in particular most recent approaches based on Machine Learning (ML) and Deep Learning techniques.
In general, these techniques require large amounts of annotated data, which is sometimes difficult to obtain in the medical domain, and especially in tumour PD-L1 essays. As explained by Dimitriou et al. \cite{overview_wsi_dl}, this issue can be partly overcome by tile-level annotations, so that the same WSI can be employed to generate multiple training image tiles. However, tile-level annotations are time consuming and can be performed only by trained medical personnel, and thus not easy to obtain in many medical contexts.

The main objective of this work is to introduce and compare automated methods for PD-L1 scoring on breast tumor WSIs, using image-level as opposed to tile-level annotations. These methods are based on extracting a representation of each WSI, both with heuristic and Deep Learning approaches, and feeding it into a downstream classifier. The models are suitable even for contexts where a small amount of images is available, which is our case. These models do not aim to replace the experience of a pathologist, but to support their decisions providing reproducible and unbiased results.

\section{Related Works}
Existing studies on automated PD-L1 detection on WSIs have explored other cancer tissues, while neglecting breast cancer. 
Most importantly, they propose and usually develop deep learning models for PD-L1 detection or Tumor Proportion Score (TPS) estimation based on strongly annotated slides. 
They are nonetheless of relevance, as WSIs stained for PD-L1 detection share similar features and is useful for estimating the goodness of our proposed model.

Convolutional Neural Networks (CNN) \cite{oshea2015cnnintro} are a common architecture in such tasks.
For example, Wu, J. et al. \cite{lung_pdl1} used this method for estimating discretized TPS in lung tumors over 2 small datasets (115 WSIs) with fine annotations, scoring over 90\% in both sensitivity and specificity.

Another popular approach for this task is using image segmentation methods to discern positive tumor areas; again, this approach needs fine segmentation annotations to be present.
For example, both works of Jianghua Wu et al. \cite{WU2022403unet} and Ziling Huang et al. \cite{HUANG2022106829unet}, use a U-Net network \cite{ronneberger2015unet} for the image segmentation phase and then calculate the TPS based on the number of segmented positive tumor cells. The first study uses  stained WSIs and obtains an accuracy of 96.24\%, while the second employs Dako clone 22C3 assay and scores 79.13\% in for TPS accuracy.
Wang et al. \cite{pdl1_breast_ring_study} present an AI model for estimating PD-L1 IC score, that, instead, employs a Linknet \cite{Chaurasia2017linknet} based model for the segmentation phase; results show a pathologists concordance score of AI predictions over 98\% for binary PD-L1 classification.

A different approach \cite{GAN_pdl1} uses Auxiliary Classifier Generative Adversarial Networks (AC-GAN)\cite{odena2017acgan} to classify tiles to be positive or negative to PD-L1 expression in a semi-supervised way, i.e. only a portion of the tile dataset is labeled. This method leads to an average overall concordance score of 87\% with TPS predictions of pathologists.

\section{Data}
\label{sec:data}

\begin{figure*}[h]
\centering
\begin{subfigure}[b]{\textwidth}
    \centering
    \includegraphics[width=1\linewidth]{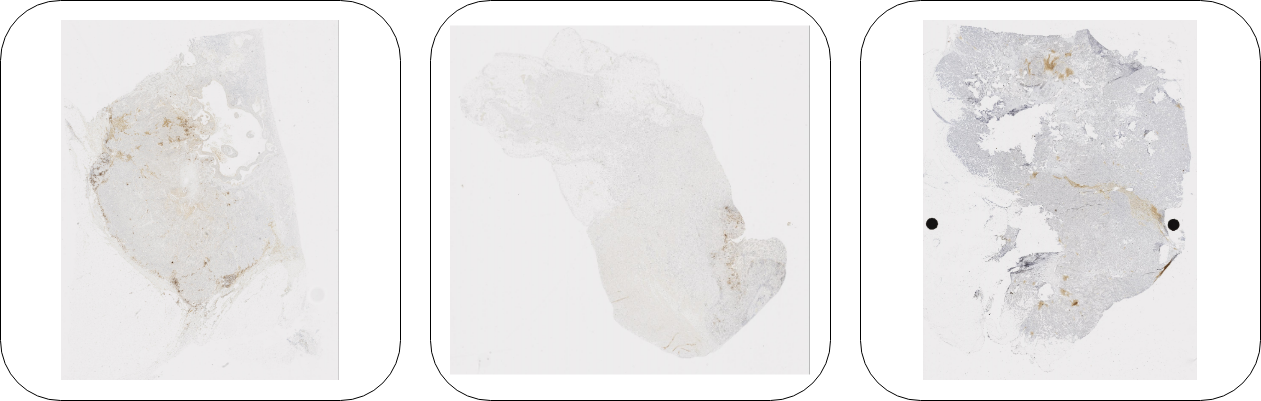}
    \label{fig:examples_wsi_int}
    \vspace{0.5cm}
\end{subfigure}
\begin{subfigure}[b]{\textwidth}
    \centering
    \includegraphics[width=1\linewidth]{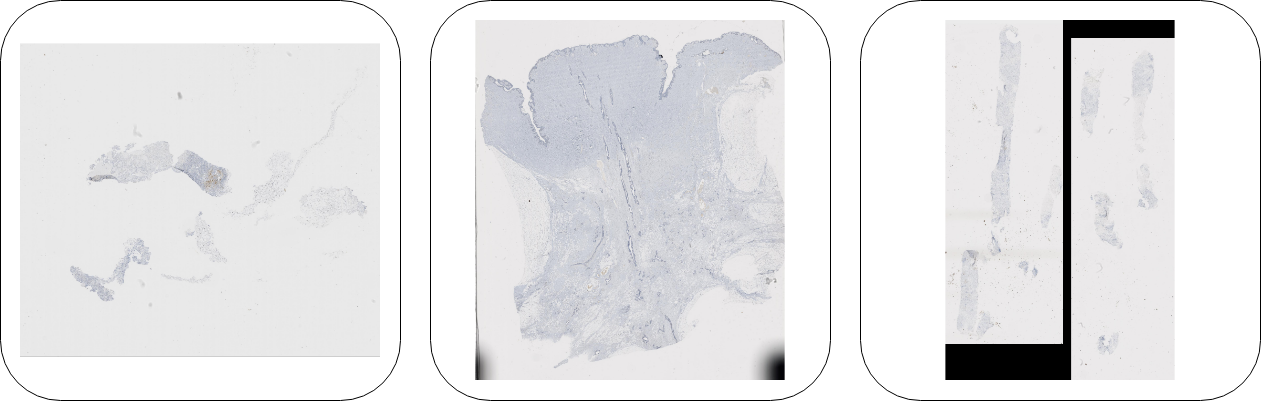}
    \label{fig:examples_wsi_ext}
\end{subfigure}
\caption{Examples of WSIs in the datasets. Top row WSIs are from the first dataset, bottom row from the second dataset (\textit{external test set}). Artifacts are easily noticeable in some of these examples.}
\label{fig:examples_wsi}
\end{figure*}

We employ two datasets of WSIs of BC tumors collected by two different institutions, all stained with the VENTANA PD-L1 (SP142) Assay. The resulting WSIs show evident brown colored stains in areas and cells which have a high PD-L1 presence which are used in our approach to identify PD-L1 \cite{ventana}. The first dataset consists of 39 WSIs, 20 positive and 19 negative to PD-L1 staining, collected at the European Institute of Oncology. 
The second dataset is composed of 25 WSIs, 21 negative and 4 positive, with similar characteristics to the first dataset, but with a possible slight domain shift caused by the different scanning equipment. These were collected at the University Hospital of the University of Pisa. 

In both datasets each slide has a full resolution in the range of approximately 50.000 to 150.000 pixels both in width and height. The images contain breast tumor regions with varying degrees of PD-L1 staining, various ROI sizes and artifacts, and have image-levele annotation as PD-L1 positive or negative. Due to WSI large resolutions, tiling is needed for both computational feasibility and for enabling exploitation of the spatial features of the WSIs. In our case we used 256x256px tiles downsampled to 64x64px.

\subsection{Manual Artifact Removal}
\label{sec:manual_art_rem}
Figure~\ref{fig:examples_wsi} shows example WSIs from the two datasets. Some images may have artefacts included during the measuring process. While our methodology can recognise some artifact types, some can be missed. In order to test our models in a more favourable setting, where the automatic tools are combined with expert knowledge, we created a replicate of our dataset where brown artifacts, that our method does not remove from the ROI, are manually removed.    
Four WSIs from the first dataset and five from the second dataset were subjected to manual artifact removal. We will test our models both with and without this manual artifact removal step.

\section{Models}
Our objective is to classify a given WSI as positive or negative to PD-L1 staining. Since the PD-L1 percentage is calculated by pathologists as a percentage only on the area of the tumor (and not on the whole WSI), the modeling pipeline is divided into two phases. The first aims at recognising the region of interest (ROI). The second aims at understanding if the tumor area is PD-L1 positive or negative. 

\begin{figure*}[h]
\centering
\includegraphics[width=1\linewidth]{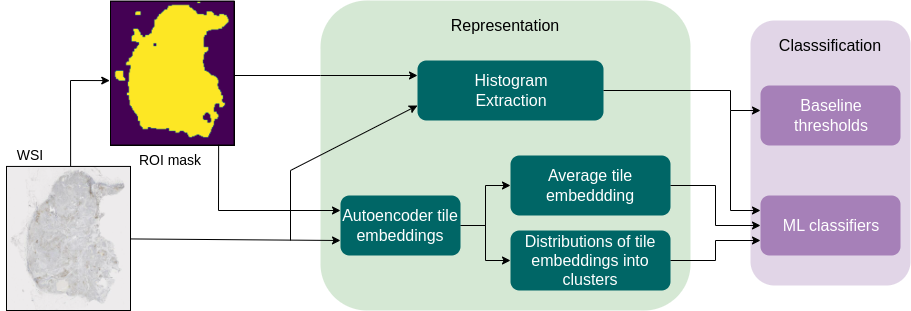}

\vspace{0.2cm}
\caption{Pipeline: ROI identification, WSI representation and final classification.}
\label{fig:pipeline_diagram}
\end{figure*}

Fig.~\ref{fig:pipeline_diagram} shows the analysis pipeline for the explored models. The initial ROI identification step is common to all models; the differences are in the condensed WSI representation step. We explore two types of image representations: (1) the histogram of the distance of pixel colors from brown and (2) embeddings extracted with an autoencoder. For the latter, we use two different types of aggregation of tile embedding to obtain WSI-level embeddings:  average embedding and distribution of embeddings into clusters.
The final classification step uses ML classifiers across all models, besides the baseline threshold approach used only on the histogram representation. The models and analysis pipeline were implemented in Python using Scikit-learn, OpenSlide and Pytorch, and the code is freely available on \url{https://www.github.com/giacomo-cgn/pdl1-wsi-classification}.

\subsection{ROI Identification}\label{sec:roi_identification}
To identify the tumor ROI we used an heuristic based on the color distance to white. We observed that the WSI areas outside the ROI have pixels of a particular shade of white, RGB[238,238,238]. The CIELAB\cite{lab_color} color space was chosen, as it is designed to approximate human vision, mimicking the analysis of a pathologist. To calculate the distance in the CIELAB space, CIEDE2000 was used \cite{ciede2000}.
The algorithm calculates the color distance from white for each pixel in a tile, to evaluate the belonging of a tile to the ROI. The algorithm is providing an approximation of the ROI, as the classification of border tiles can be noisy.

In order to be able to recognise artifacts that present as dark (almost black) patches, we consider the pixels that have a distance from white very different from other pixels from the same WSI, i.e. outliers. We consider outliers all distances that are more that 3 standard deviations away from the mean. Tiles with a large number of outlier pixels are considered outliers.

More specifically the algorithm works as follows. For each tile, we compute the distance from white for all pixels. If over 80\% of the tile pixels are outliers,  
we mark the tile as an artifact and set its ROI mask value to 0. Otherwise we compute the fraction of pixels which have a distance from white lower than 5 (obtained through preliminary trials), resulting in a float mask for the WSI (one fraction per tile).
We then binarize the float mask with a simple thresholding approach, i.e. tiles with a fraction lower than a threshold $F_\text{ROI}$ are considered inside the ROI. 
Subsequently, to obtain a smoother and more contiguous ROI, the morphological operations of closing and opening\cite{opening_closing} are applied to the binary mask in this order.
The resulting  binary matrix is the final ROI mask used in the next steps for PD-L1 detection.

\subsection{PD-L1 Classification through Colour Distance Histograms}\label{sec:histogram}
The regions of the WSI with high concentration of PD-L1 staining can be algorithmically identified using their brown colouring, in a similar way to the ROI identification (Section \ref{sec:roi_identification}). The pixels belonging to a PD-L1 stained areas have relatively high color variance, thus the shade of brown to be used as base color was extracted averaging a set of 16 different PD-L1 pixels; the resulting color RGB[117.3, 88.9, 67.3] is referred as \texttt{base\_brown}.

\begin{figure*}[h]
\centering
\includegraphics[width=1\linewidth]{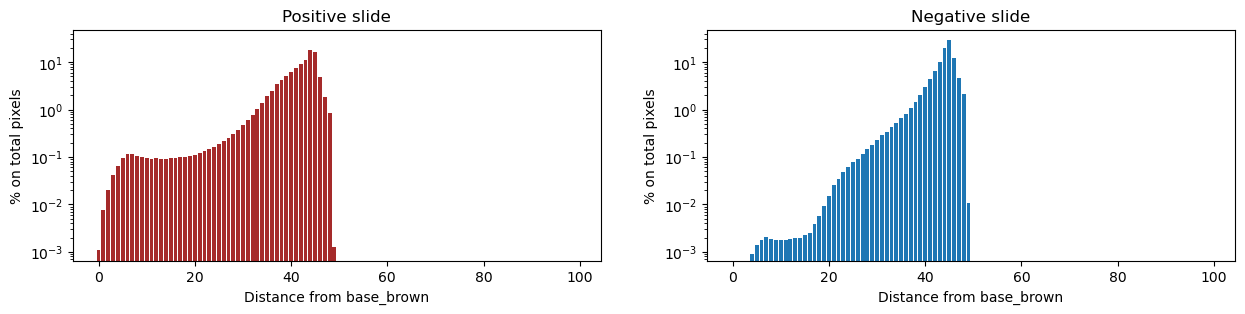}
\caption{Histograms with 100 bins in logarithmic scale of a positive and negative WSI. We observe that for the positive slide the fraction of pixels in the lowest part of the distribution is higher. }
\label{fig:hist_brown}
\end{figure*}

We therefore compute a representation of the WSI by using the histogram of the distance to \texttt{base\_brown}, applied only to the tiles in the ROI, that we then log-normalise. We use histograms with 100 bins.
Two examples of the histograms obtained for a negative and a positive WSI are shown in Fig.~\ref{fig:hist_brown}.

We develop two different approaches for the classification of the WSI based on the histogram representation, which are explained in the following subsections:
\begin{itemize}
    \item The \textbf{Baseline Histogram Model}.
    \item The \textbf{ML Histogram Models}.
\end{itemize}

\subsubsection{Baseline Histogram Model}
The \textit{Baseline Histogram Model} uses 2 thresholds in order to predict the positivity of a slide based on the previously calculated representation (histogram). 
This heuristic approach is justified by the fact that the histogram in log-space of positive slides has a bimodal distribution, as seen in Fig. \ref{fig:hist_brown}; the first peak corresponds to the significant number of pixels close to \texttt{base\_brown}.
The first threshold is $t_\text{bin}$ and it roughly denotes up to which bin a color is considered close to \texttt{base\_brown}. 
Thus, the ratio of of pixels close to \texttt{base\_brown} can be defined as $r$.
The second threshold $t_\text{cls}$ is applied on $r$ and defines over which value of $r$ a slide is classified as positive. Selected thresholds are the ones achieving the best accuracy on the training set, using an extensive grid search. 
Algorithm \ref{alg:baseline} illustrates in detail the process of classifying a WSI according to the \textit{Baseline Histogram Model}.
We recall that, for this classifier, a validation set is not needed as the training process itself is the choice of the thresholds.

\begin{algorithm}
\caption{Algorithm for the histogram calculation and \textbf{histogram baseline model}.}\label{alg:baseline}
    \begin{algorithmic}
    \State \textbf{Given:} Bin Threshold $t_\text{bin}$
    \State \textbf{Given:} Classification Threshold $t_\text{cls}$
    \State \textbf{Given:} \textit{base\_brown}
    \State \textbf{Given:} \textit{num\_bins}
    \\
    \\
    /* \texttt{ Histogram calculation } */
    \State $\textit{distances\_list} \gets [ \ ]$
    \For{pixel $p$ in W}
        \State $\textit{distance} \gets \text{CIEDE2000}(p, \textit{base\_brown})$
        \State $\textit{distances\_list}.\text{append}(\textit{distance})$
    \EndFor

    \State $\textit{hist} \gets \text{calculate\_histogram}(\textit{distances\_list}, \textit{num\_bins})$
    \\
    \\
    /* \texttt{ Baseline Histogram Model classification } */
    \State $r = \dfrac{\sum_{i=0}^{t_\text{bin}}{\textit{hist}[i]}} {\sum_{i=0}^{\textit{num\_bins}}{\textit{hist}[i]}}$
    \\
    \If{$r > t_\text{cls}$}
        \State $W$ is positive
    \Else
        \State $W$ is negative
    \EndIf
\end{algorithmic}
\end{algorithm}

\subsubsection{ML Histogram Models}\label{sec:method_histogram_ml}
In the second approach for histogram classification, the WSI representations are used as input for training ML models, therefore, each model has 100 input features. 
The possible advantage of using this method is that the ML model is able to learn the histogram shape that maximises accuracy, replacing the need for less flexible thresholds.
The chosen ML models are Random Forest (RF) and Support Vector Machine (SVM). Hyperparameter tuning on each model is done by an exhaustive grid search, through 6-fold cross validation.
After the best set of hyperparameters for each model has been found, that model is then retrained on the whole training set with those hyperparameters.

\subsection{PD-L1 Classification through Autoencoder Embeddings}\label{sec:autoencoder_model}
In this second class of models, we used a convolutional autoencoder (CAE) to extract representations for tiles, that are then aggregated to obtain WSI representations. 
The CAE is a  convolutional neural network architecture that is used to learn a compact representation, or embedding, of image data \cite{convolutional_ae}.

The network is trained in a self-supervised manner, i.e. it aims to reproduce the images provided as input. Specifically, the network consists of two sequential components: the encoder, which learns to extract meaningful features from the input image, and the decoder, which learns to reconstruct the original image from these features. The features extracted by the encoder part become our representation of the image.

The robustness of tile-level embedding extraction through an autoencoder in WSI analysis has already been proven by David Tellez et al. \cite{autoencoder_camelyon} who successfully used it to extract patch embeddings from the CAMELYON16 challenge WSI dataset.
We employed a tile-level, instead of a WSI-level CAE, in order to reduce the size of the input. In this way, we also had a significantly larger amount of training samples (each WSI contributed a large set of tiles), with evident benefits in term of generalization capabilities.

\begin{figure*}[h]
\centering
\includegraphics[width=1.05\textwidth]{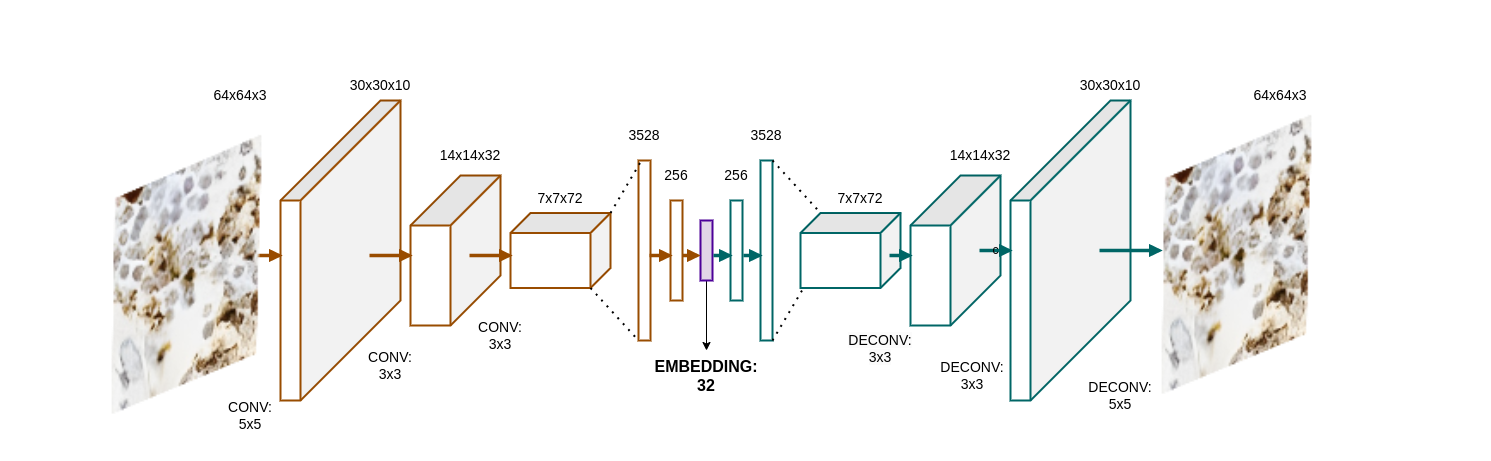}
\caption{The figure represents the CAE structure. The encoder uses convolutions with max pooling, while the decoder uses padding. Batch normalization and small convolutional and deconvolutional kernels (3x3, 5x5) are used. The final embedding (purple) is extracted after the 2 fully connected encoder layers. }
\label{fig:ae_diagram}
\end{figure*}

In our case, we built a CAE with 3 convolutional layers and 2 fully connected (FC) layers for the encoder and again 3 deconvolutional layers and 2 FC layers for the decoder (Fig. \ref{fig:ae_diagram}). The CAE takes 64x64px tiles  belonging to the previously identified ROI as an input. We train on the complete CAE architecture and use only the encoder at prediction time to extract the 32-dimensional tile embeddings generated after the encoder FC layers. Training was executed using a learning rate of 0.001 and 20 training epochs were employed for reaching convergence.

The resulting set of 32-dimensional tile embeddings for each WSI are aggregated to obtain one WSI embedding. We use two types of aggregation, both spatially invariant, based on the assumption that only local visual information is relevant for PD-L1 detection:
the position of PD-L1 staining inside the WSI does not influence the final PD-L1 WSI score.

The two kinds of aggregation differentiates the two autoencoder embeddings based models, that are:
\begin{itemize}
    \item \textbf{Average Tile Embeddings Model}, uses as an aggregation strategy the average of tile-level embeddings, resulting in one 32-dimensional embedding vector for each WSI.
    \item \textbf{Clustered Tile Embeddings Model} uses clustering to gather the tile embeddings into similar groups (tiles with similar features), employing K-Means. 
    Clusters are defined on all the tile embeddings extracted from the training set, i.e. we fix the cluster centroids to those from the training data. 
    To obtain more cohesive clusters at training time, for each WSI, only the closest $t_\text{op}$ percentile of tiles, ordered by distance from their centroid, is kept assigned to their cluster, where $t_\text{op}$ is the outlier percentile threshold.
    Instead, at evaluation time, we keep assigned to each cluster only tiles being closer to the center than the furthest training tile assigned to that cluster. 
    The remaining distant tiles are marked as outliers and assigned to a general fictitious cluster for outliers labeled as \textit{$-1$}.
    Finally, we use the fraction of tiles in each group as the final representation (WSI embedding). Thus, the compact WSI representation is the \textit{$K+1$}-dimensional distribution of tiles into clusters (\textit{K} clusters + \textit{$1$} outlier label).
    \end{itemize}

Both WSI aggregated embeddings are then passed as input into two ML classifiers, SVM and RF, whose output is the positive or negative classification for the entire WSI to PD-L1 staining.
Hyperparameters selection for these 2 models is done by an extensive grid search, through cross validation on the training set, similar to section \ref{sec:method_histogram_ml}.

\subsection{Experimental setting}

Given that, as explained in Section \ref{sec:data}, we have 2 differently sourced dataset, we employ two distinct experimental training and testing scenarios:
\begin{itemize}
    \item In the first one, referred as \textit{combined datasets}, the two datasets are concatenated; training and test set are extracted from the combination and thus contain WSIs coming from both datasets. This configuration aims to evaluate the capabilities of our models when exposed to more diverse samples in the training phase.
    \item In the second one, referred as \textit{separate datasets}, we compose the training set with slides coming only from the first dataset, while slides from the second dataset are reserved for testing only. In this scenario, we have a total of 2 test sets: the \textit{internal} one, composed from the remaining samples form the first dataset, and the \textit{external} one which is composed by all the samples from the second dataset.
This testing configurations aims to evaluate the generalization capabilities of our models.
\end{itemize}

Additionally, we also tested all of our models \underline{without} the manual brown artifact removal step, described in Section \ref{sec:manual_art_rem}, in order to test the resilience of our methods to the presence of artifacts that could be easily mistaken for PD-L1 staining.

For the \textit{Clustered Tile Embeddings Model}, we tested various values for $K$ in K-Means and for the outlier percentile threshold $t_\text{op}$; based on performances on validation data, we used the model employing 256 clusters and $t_\text{op} = 90$.

\section{Results}

\subsection{ROI identification}

The ROI identification process was evaluated qualitatively, and was deemed to return a good approximation of the tumor region for most WSIs. In some slides with a particularly light tumor area, relatively close to white, the algorithm is not capable of identifying clearly a ROI. In these, changes in the ROI threshold greatly vary the size of the identified ROI.

\begin{figure*}[!htbp]
    \centering
    \begin{subfigure}[b]{0.33\textwidth}
        \centering
        \includegraphics[width=\textwidth]{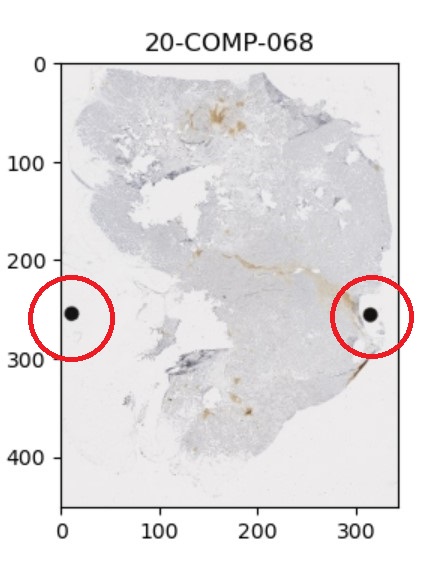}
        \caption{\small Original WSI}  
        
    \end{subfigure}
    \hfill
    \begin{subfigure}[b]{0.3\textwidth}  
        \centering 
        \includegraphics[width=\textwidth]{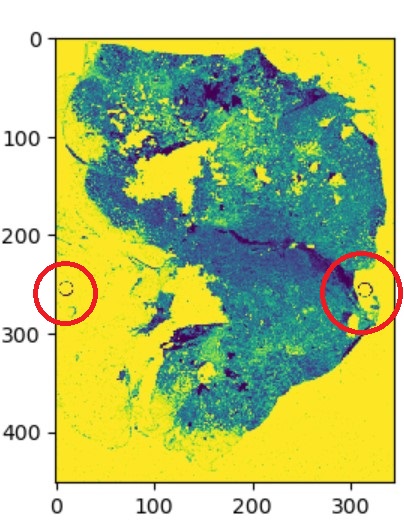}
        \caption{\small ROI float mask after dark artifact reduction has been applied}
    \end{subfigure}
    \hfill
    \begin{subfigure}[b]{0.3\textwidth}   
        \centering 
        \includegraphics[width=\textwidth]{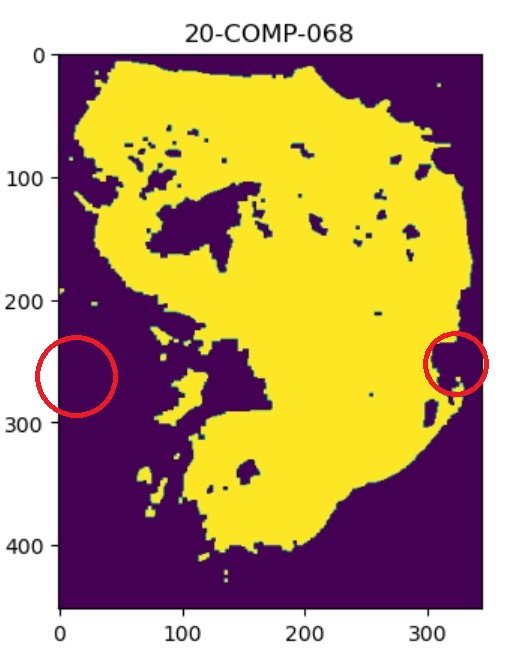}
        \caption{\small ROI map with threshold 0.85 from ROI identification with dark artifacts reduction.}
    \end{subfigure}
    \caption
    {WSI with 2 dark artifacts and confrontation with resulting ROI identification processes with artifacts reduction}
    \label{fig:result_roi_norm}
\end{figure*}

The preliminary exclusion of artifacts in our ROI identification pipeline is effective mostly for the darker and most uniform ones. On the other hand, the effectiveness of this method is limited for lighter coloured and faded artifacts, as their tiles rarely have most of the pixel color distance from white above the found \textit{maximum} for the slide. A successful exclusion of an artifact can be observed in the slide included in Figure \ref{fig:result_roi_norm}, in which 2 artifacts are eliminated from the ROI map.

\subsection{PD-L1 Classification}

\begin{figure*}[b]
\centering
\includegraphics[width=1\linewidth]{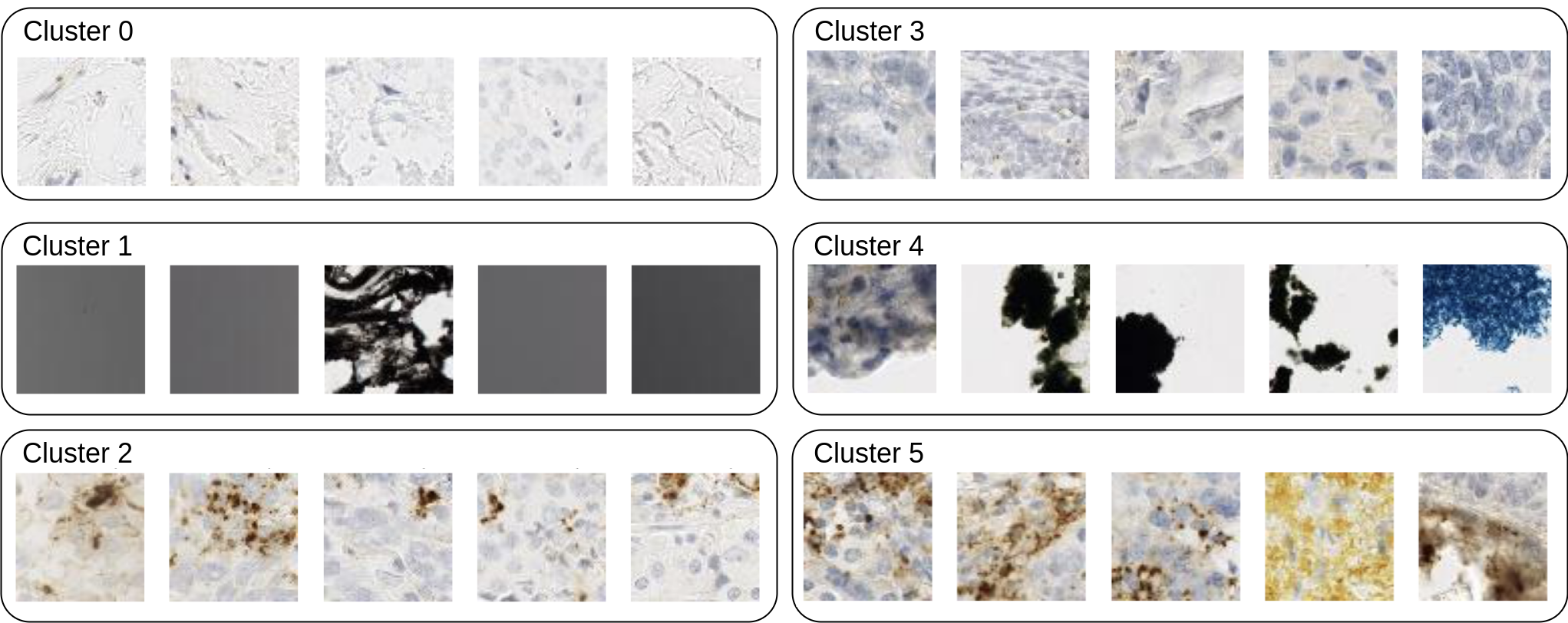}
\caption{Samples of tiles clustered by K-Means using their autoencoder embeddings. It is significant to note that each clusters seems to differentiate artifacts and different kinds of tumoral tissue present in the tiles.}
\label{fig:cluster_tile_embed_ae}
\end{figure*}

A first evaluation concentrates on the tile embedding clustering necessary for the \emph{Clustered Tile Embeddings Model. } Figure~\ref{fig:cluster_tile_embed_ae} shows a sample of images in six different clusters. We note that, qualitatively, clusters contain similar tiles. Some of the clusters represent PD-L1 positive areas, some are negative areas, and some are artefacts. This suggests that the Deep Learning model could be able to exclude artefacts without manual intervention. 

\begin{table}[h!]
    \caption{Results of models trained and tested \textbf{\textit{with} the manual artifact removal step}.}

    \begin{tabular}{ p{3.0cm}|p{1.0cm}|p{5.0cm} }
     \multicolumn{2}{c|}{\textbf{Model}} & \textbf{Combined} test set \\
     \toprule
     \multicolumn{2}{c|}{Baseline Histogram} & \textbf{91.67\% (tp:4 fn:1 tn:7 fp:0)}\\
     \midrule
     \multirow{2}{4em}{ML Histogram}
                 & SVM & 50.00\% (tp:1 fn:4 tn:5 fp:2)\\
                 & RF & \textbf{91.67\% (tp:4 fn:1 tn:7 fp:0)}\\
     \midrule
      \multirow{2}{8em}{Average Tile Embeddings}
                 & SVM  & 66.67\% (tp:3 fn:2 tn:5 fp:2)\\
                 & RF  & 75.00\% (tp:3 fn:2 tn:6 fp:1)\\
    \midrule
     \multirow{2}{8em}{Clustered Tile Embeddings}
                 & SVM & 83.33\% (tp:4 fn:1 tn:6 fp:1)\\
                 & RF  & 75.00\% (tp:3 fn:2 tn:6 fp:1)\\
    \end{tabular}

    \vspace{0.3cm}
     
    \centering

    \begin{tabular}{ p{1.8cm}|p{0.7cm}|p{4.4cm}|p{4.4cm} }
     \multicolumn{2}{c|}{\textbf{Model}} & Separated test sets - \textbf{Internal} & Separated test sets - \textbf{External} \\
     \toprule
     \multicolumn{2}{c|}{Baseline Histogram} & 66.67\% (tp:3 fn:0 tn:1 fp:2) & 76.00\% (tp:3 fn:1 tn:16 fp:5)\\
     \midrule
     \multirow{2}{4em}{ML Histogram}
                 & SVM & 66.67\% (tp:3 fn:0 tn:1 fp:2) & 68.00\% (tp:2 fn:2 tn:15 fp:6)\\
                 & RF & 50.00\% (tp:3 fn:0 tn:0 fp:3) & 80.00\% (tp:2 fn:2 tn:18 fp:3)\\
     \midrule
      \multirow{2}{8em}{Average Tile Embeddings}
                 & SVM & 83.33\% (tp:2 fn:1 tn:3 fp:0) & 88.00\% (tp:1 fn:3 tn:21 fp:0)\\
                 & RF & \textbf{100.00\% (tp:3 fn:0 tn:3 fp:0)} & \textbf{88.00\% (tp:1 fn:3 tn:21 fp:0)}\\
    \midrule
     \multirow{2}{8em}{Clustered Tile Embeddings}
                 & SVM & \textbf{100.00\% (tp:3 fn:0 tn:3 fp:0)} & 80.00\% (tp:0 fn:4 tn:20 fp:1)\\
                 & RF  & 83.33\% (tp:3 fn:0 tn:2 fp:1) & \textbf{88.00\% (tp:1 fn:3 tn:21 fp:0)}\\
    \end{tabular}

    \footnotetext{Reported scores are in the form: \textit{accuracy\% (true positives, false negatives, true negatives, false positives)}.}
   
    \label{tab:final_results}
\end{table}

The final PD-L1 classification performance in the different dataset combination settings, on data with manual artifact reduction, is reported in Table \ref{tab:final_results}. We report accuracy and also true/false positives/negatives, for a complete evaluation of performance on an imbalanced dataset.

When combining the datasets, two models based on the histogram features (RF and the baseline) reach the optimal score. The Clustered Tile Embeddings Models are not far in term of accuracy.
It is crucial to note that, in presence of diverse and cleaned enough training data, the strong heuristic assumptions of the histogram models, especially the baseline, prove to be enough for a reliable classifier.

For separated datasets, we can easily see that the autoencoder embeddings based models achieve significantly higher scores, especially in the internal test set, in which both the RF on the Average Embeddings and the SVM on the Clustered Embeddings achieve a flawless score. Nonetheless, they encounter greater challenges in recognizing the few positive WSIs in the external test set (3/4 False Negatives). This shows a tendency towards negative classification in presence of this domain shift, probably due to some form of overfitting on the internal training dataset.
In contrast, the histogram based models show a tendency towards positive classification, with a higher False Positives count, in both internal and external test sets. But, for this same reason, they retain an higher accuracy on individuating positives in the unbalanced external test set, thus implying less overfitting on the internal training dataset.

\begin{table}[h]
    \caption{Results of models trained and tested \textbf{\textit{without} the manual artifact removal step}. }

    \centering
    \begin{tabular}{ p{3.0cm}|p{1.0cm}|p{5.0cm}  }
     \multicolumn{2}{c|}{\textbf{Model}} & \textbf{Test}\\
     \toprule
     \multicolumn{2}{c|}{Baseline Histogram} 
                    & 83.33\% (tp:3 fn:2 tn:7 fp:0)\\
     \midrule
     \multirow{2}{4em}{ML Histogram}
                 & SVM & 66.67\% (tp:3 fn:2 tn:5 fp:2) \\
                 & RF & 66.67\% (tp:1 fn:4 tn:7 fp:0) \\
     \midrule
      \multirow{2}{8em}{Average Tile Embeddings}
                & SVM            & \textbf{91.67\% (tp:4 fn:1 tn:7 fp:0)} \\
                & RF             & 83.33\% (tp:3 fn:2 tn:7 fp:0) \\
     \midrule
     \multirow{2}{8em}{Clustered Tile Embeddings}
                 & SVM          & 66.67\% (tp:2 fn:3 tn:6 fp:1) \\
                 & RF           & 83.33\% (tp:3 fn:2 tn:7 fp:0) \\
    \end{tabular}

    \vspace{0.3cm}

    \begin{tabular}{ p{1.8cm}|p{0.7cm}|p{4.4cm}|p{4.4cm}  }
     \multicolumn{2}{c|}{\textbf{Model}} & Separated test sets - \textbf{Internal} & Separated test sets - \textbf{External} \\
     \toprule
    \multicolumn{2}{c|}{Baseline Histogram} 
                   & 50.00\% (tp:3 fn:0 tn:0 fp:3) & 76.00\% (tp:3 fn:1 tn:16 fp:5)\\
     \hline
     \multirow{2}{4em}{ML Histogram}
                 & SVM & 50.00\% (tp:3 fn:0 tn:0 fp:3) & 64.00\% (tp:2 fn:2 tn:14 fp:7)\\
                 & RF & 50.00\% (tp:3 fn:0 tn:0 fp:3) & 76.00\% (tp:2 fn:2 tn:17 fp:4)\\
     \midrule
      \multirow{2}{8em}{Average Tile Embeddings}
                & SVM            & 83.33\% (tp:2 fn:1 tn:3 fp:0) & 64.00\% (tp:1 fn:3 tn:15 fp:6) \\
                & RF             & 83.33\% (tp:2 fn:1 tn:3 fp:0) & 48.00\% (tp:2 fn:2 tn:10 fp:11)\\
    \midrule
     \multirow{2}{8em}{Clustered Tile Embeddings}
                 & SVM          & 50.00\% (tp:0 fn:3 tn:3 fp:0) & 80.00\% (tp:0 fn:4 tn:20 fp:1) \\
                 & RF           & \textbf{100.00\% (tp:3 fn:0 tn:3 fp:0)} & \textbf{88.00\% (tp:1 fn:3 tn:21 fp:0)} \\
    \end{tabular}

    \footnotetext{Reported scores are in the form: \textit{accuracy\% (true positives, false negatives, true negatives, false positives)}.}
   
    \label{tab:results_artifacts}
\end{table}

Table \ref{tab:results_artifacts} shows the outcomes of experiments without brown artifact removal.
In the case of combined datasets, leaving the artifacts untouched appears to increase performance in most CAE tile embeddings models, compared to those trained on data without artifacts (Table \ref{tab:final_results}). The SVM with Average Embeddings model displays best performance. This could indicate that artifact-introduced variability allows to build a better embeddings, which also enables the management of artifacts without manual intervention. 
In the case of separated datasets, the performance is similar for all CAE tile embeddings models when introducing artifacts, except for the SVM on the Clustered Embeddings Model, which has a sudden decrease to 50\% accuracy. The best performance is that of the RF with Clustered Tile Embeddings.
In the case of histogram based models, as expected, the presence of brown artifacts drastically reduces performance for the classifiers in both the separated and the combined test sets, especially in the case of the SVM. This shows the superiority of deep learning models in a setting without manual pre-processing. 
Similar to the case without artifacts, we observe a tendency of false negative classification in the external test set for the CAE tile embeddings models, while histogram based models seems slightly more resilient in correct positives classification, when a domain shift appears in the test set. However, histogram-based models also have a lot of false positives, reducing their overall accuracy.

\section{Conclusions}
The objective of this work was to create a model capable of classifying PD-L1 positivity on a WSI using weakly-labelled data. We compared two types of representations for images, fed into downstream ML classifiers: a representation based on the colour distance and representations that aggregate CAE tile embeddings. 

The study was performed on two WSI datasets from different clinical centers, employing two testing configurations: training on one dataset and testing the other, versus combining the datasets. 
We repeated training and testing of the models for both the presence and absence of a human-in-the-loop scenario for brown artifact removal.

In presence of the artifact removal process, for the combined testing configuration the histogram models seem to be superior (baseline and SVM), while for the separated testing configuration the CAE tile embeddings models take the advantage, in particular the RF with Average Tile Embeddings Model, as it reaches the optimal score in both internal and external test sets.
Instead, leaving artifacts untouched, the strong assumptions on color histograms do not hold as well and the best scores are achieved by CAE tile embeddings models.

All in all,  with a training set with limited variance and when a possible domain shift from training to testing is to be expected, the CAE tile embeddings models with a RF classifier proves to be the most robust model.
Instead, when the data are more similar between training and testing  and brown artifacts have been thoroughly hand removed, simpler histogram based models are reliable enough.
These results confirm the paper's primary aim: proving the reliability of models based on WSI-level labels, as an alternative to the more common and expensive paradigm of tile-level labels.

Furthermore, our modelling pipeline is modular and open for future improvements and adaptations.
ROI identification and PD-L1 scoring are separate steps, enabling, for example, the introduction of more advanced approaches for ROI identification or of other ML models for scoring.
Future work could improve the embedding extraction and aggregation process, either with network architectures alternative to the current autoencoder or with other aggregation procedures.
Stronger assumptions could be made about the spatial invariance of the tiles when training the CAE, for example by using smaller tiles or by enforcing rotation invariance to augment the training dataset. Transfer learning could also be employed, by taking advantage of other WSI datasets not specific to PD-L1 or BC.

Importantly, our method does not aim to perform scoring alone, but to be a tool to support pathologists in their decision making.  We do not plan to exclude the manual artifact removal step, but we plan to develop a QuPath \cite{Bankhead2017} plugin that allows for interaction between the pathologist and our model, especially in the definition of the ROI and in artifact identification.

\bmhead{Acknowledgements}
This work has been partially funded by the SPARK initiative at the University of Pisa, through the project "DROP: Digital Research in Oncologic Pathology".

\bibliography{refs}

\end{document}